\documentclass[twocolumn]{aastex63}
\usepackage{xcolor}
\usepackage{mathtools}
\usepackage[caption = false]{subfig}

\newcommand{\msun}{M$_{\odot}~$} 
\newcommand{\msune}{M$_{\odot}$}

\newcommand{\h}{$^{\rm h}$}
\newcommand{\m}{$^{\rm m}$}
\newcommand{\s}{$^{\rm s}~$}

\newcommand{\feh}{[Fe/H]}
\newcommand{\ofe}{[O/Fe]}
\newcommand{\afe}{[$\alpha$/Fe]}

\shorttitle{Exploring the Evolution of Massive Clumps}
\shortauthors{Garver et al.}

\begin{document}

\title{
Exploring the Evolution of Massive Clumps in Simulations that Reproduce the \\ Observed Milky Way $\alpha$-element Abundance Bimodality}
\correspondingauthor{Bethany R. Garver}
\email{bethanygarver@montana.edu}

\author[0000-0001-6013-9125]{Bethany R. Garver}
\affiliation{Department of Physics, Montana State University, P.O. Box 173840, Bozeman, MT 59717-3840}

\author[0000-0002-1793-3689]{David L. Nidever}
\affiliation{Department of Physics, Montana State University, P.O. Box 173840, Bozeman, MT 59717-3840}

\author[0000-0001-7902-0116]{Victor P. Debattista}
\affiliation{Jeremiah  Horrocks  Institute,  University  of  Central  Lancashire, Preston, PR1 2HE, UK}

\author[0000-0002-0740-1507]{Leandro {Beraldo e Silva}}
\affiliation{Jeremiah  Horrocks  Institute,  University  of  Central  Lancashire, Preston, PR1 2HE, UK}
\affiliation{Department of Astronomy, University of Michigan, 1085 S. University Ave., Ann Arbor, MI 48109, USA}

\author[0000-0002-3343-6615]{Tigran Khachaturyants}
\affiliation{Jeremiah  Horrocks  Institute,  University  of  Central  Lancashire, Preston, PR1 2HE, UK}

\begin{abstract}

The Milky Way stellar disk has both a thin and a thick component. The thin disk is composed mostly of younger stars ($\lesssim$8 Gyr) with a lower abundance of $\alpha$ elements, while the thick disk contains predominantly older stars ($\gtrsim$8--12 Gyr) with a higher $\alpha$ abundance, giving rise to an $\alpha$-bimodality most prominent at intermediate metallicities. A proposed explanation for the bimodality is an episode of clumpy star formation, where high-$\alpha$ stars form in massive clumps that appear in the first few Gyrs of the Milky Way's evolution, while low-$\alpha$ stars form throughout the disk and over a longer time span. To better understand the evolution of clumps, we track them and their constituent stars in two clumpy Milky Way simulations that reproduce the $\alpha$-abundance bimodality, one with 10\% and the other with 20\% supernova feedback efficiency. We investigate the paths that these clumps take in the chemical space (\ofe--\feh) as well as their mass, star formation rate (SFR), formation location, lifetime, and merger history. The clumps in the simulation with lower feedback last longer on average, with several lasting hundreds of Myr. Some of the clumps do not reach high-$\alpha$, but the ones that do on average had a higher SFR, longer lifetime, greater mass, and form closer to the galactic center than the ones that do not. Most clumps that reach high-$\alpha$ merge with others and eventually spiral into the galactic center, but shed stars along the way to form most of the thick disk component.

\end{abstract}


\section{Introduction} \label{sec:intro}

The Milky Way (MW) disk is comprised of two populations of stars, the thick and thin disks \citep{Yoshii1982,Gilmore1983}. The thick disk is composed of mostly older stars with higher abundances of $\alpha$ elements, such as oxygen and magnesium, relative to iron abundance, while the thin disk contains mostly younger stars that have lower, roughly solar, $\alpha$-abundances \citep[e.g.][]{Fuhrmann1998,Haywood2013,Bensby2014}. There is a pronounced $\alpha$-abundance bimodality created by these two populations which is most prominent at intermediate metallicities and near the solar radius \citep[e.g.][]{Nidever2014,Hayden2015}. While there is overlap between the chemical (defined by $\alpha$-abundance) and geometrical (defined by location and thickness) thin and thick disks, the chemical thick disk stars are older than $\sim$9 Gyr \citep{Xiang_Rix_2022} and concentrated toward the center of the galaxy \citep{Bensby2011,Minchev2015}, while stars in the geometric thick disk are more radially extended and have ages of 5-9 Gyr and a radial age gradient with younger stars in the outer disk \citep{Martig2016}. The chemical thick disk also has a shorter scale length, $\sim$2 kpc, than the geometric thick disk, $\sim$3.6 kpc \citep{Juric2008, Bovy2012}.

Several possible scenarios for the origin of the separation of the chemical thin and thick disks have been proposed. The ``two-infall" model \citep{Chiappini1997,Chiappini2009,Spitoni2019} suggests that the thick disk formed from an initial gas infall on a timescale of $\sim$0.4 Gyr. The initial infall had a high star formation rate (SFR), which caused the stars formed to have a high-\afe. A second infall of low-metallicity gas occurred later, on a timescale of several Gyr, to form the thin disk. In this model, the two disks are formed by specific events, that would likely not happen in most galaxies. However, geometric thick and thin disks have been found in several other disk galaxies \citep{Yoachim2006}, and hints of a chemical bi-modality have been found in the nearby Milky Way-like galaxy UGC~10738 \citep{Scott2021}. 

Another proposed origin for the $\alpha$-bimodality is the ``superposition" model \citep{Schonrich2009}. In this model, the high-$\alpha$ stars formed first, when the metallicity of the galaxy was very low, and then as the metallicity increased, the relative $\alpha$-abundance decreased, so the newer stars formed on the low-$\alpha$ track. The thick disk was made of older stars that formed nearer the Galactic center and spread through the MW via radial migration, and the thin disk stars naturally formed later along tracks corresponding to different birth radii. The thin disk stars also migrated through the Galaxy, causing the spread in metallicities in the local thin disk. \citet{Sharma2020} used a similar model but with improvements for chemical enrichment and velocity dispersion, and were able to reproduce the bimodality for different regions of the MW. However simulations of this scenario have failed to reproduce the bimodality of the MW, instead predicting a significant number of stars in between the high-$\alpha$ and low-$\alpha$ tracks \citep{Loebman2011}. \citet{Sharma2020} used a model with many free parameters that is fit to the data, rather than a self-consistent chemical evolution model. \citet{Johnson2021} attempted to fix this. They first used a hydrodynamical MW simulation to model radial migration without free parameters. Then they input the radial migration into another simulation that models the MW as a series of concentric rings with stellar populations that are allowed to move between rings. However, they also were not able to reproduce the $\alpha$-abundance bimodality in their simulations.

In another scenario, the outer MW had a low-metallicity, low-$\alpha$ disk that was once highly inclined relative to the rest of the galaxy \citep{Renaud2021b}. In this scenario, the low metallicity low-$\alpha$ stars formed on the inclined disk in the outer galaxy, while the high metallicity low-$\alpha$ stars formed on the main disk along with the high-$\alpha$ stars. The tilted disk eventually aligned with the rest of the galaxy. As a result, the low-$\alpha$ stars that formed on the tilted disk had halo-like kinematics because of where they formed.

Yet another possibility is that the thick disk was formed through mergers with gas-rich dwarf galaxies. The gas-rich mergers cause a high star formation rate, with stars enriched in $\alpha$ elements, forming the thick disk, while the thin disk forms later \citep{Brook2004,Brook2005}. Simulations have shown that mergers with small, gas rich galaxies early in the MW's formation are capable of forming a thick disk with high $\alpha$-abundance and high velocity dispersion similar to what is observed in the MW \citep{Brook2007,Grand2018}.

Several other studies have used simulations to model the chemical evolution of the MW and were able to reproduce the $\alpha$-bimodality \citep[e.g.][]{Kobayashi2011,Haywood2015,Haywood2016,Vincenzo2020}. \citet{Haywood2015,Haywood2016} used a closed box model for different star formation histories of the MW, and were able to reproduce the $\alpha$-bimodality with models that have a lull in star formation approximately 8 Gyr ago between the episodes of high-$\alpha$ and low-$\alpha$ star formation.

\citet{Clarke2019} proposed that the high-$\alpha$ stars formed in massive clumps that developed early in the MW's evolution. These clumps formed from gravitational instabilities in the gas disk of the galaxy \citep{Noguchi1999,Immeli2004b,Inoue2018}. The high SFR in the clumps cause them to self-enrich in $\alpha$ elements, while low-$\alpha$ stars form throughout the lower-SFR disk. The clumps also vertically scatter stars to form a geometric thick disk \citep{Bournaud2009}. These types of clumps are visible in high redshift galaxies that have not yet formed a bulge \citep{Elmegreen2005,Elmegreen2007} and their sizes are on the order of 100 pc \citep{Livermore2012}. The masses and lifetimes of clumps in high redshift galaxies are not well constrained, but the masses are likely less than $10^9$ \msun and probably in the range of $3\times 10^6$ to $5 \times 10^8$ \msun \citep{Benincasa2019}, and the lifetimes are probably less than 1 Gyr \citep{Claeyssens2023}. They appear in 60\% of low-mass galaxies at redshift between 3 and 0.5, and 55\% of intermediate and high mass galaxies at $z \sim 3$ \citep{Guo2015}. The fraction of galaxies with clumps increases with decreasing mass or increasing redshift, and the fraction of the galactic mass in clumps increases with increasing redshift \citep{Sok2021}. Other studies have simulated clumpy high-redshift galaxies. \citet{Vincenzo2019} found gas rich clumps that had high SFR along their simulated galaxy's spiral arms as the arms formed. The number and size of these clumps are greatest at $z\sim 2$. \citet{Inoue2019} tested how different models of the interstellar medium affected the clumpiness of the simulated galaxy, and found that the model with a softer equation of state resulted in significantly more clumps than the other models, without much effect on the global properties of the galaxy. \citet{Renaud2021} looked at the effect of gas fraction on the structure of the galaxy, and found that galaxies that formed with a higher gas fraction were clumpy, while those that formed with a lower gas fraction had more of a spiral structure.

\citet{Clarke2019} used a clumpy simulation that was able to reproduce an $\alpha$-abundance bimodality similar to that of the MW. The clumps in the simulation appear to have similar properties to clumps observed at high-$z$. \citet{Clarke2019} created
mock images of this simulation at $z=2$ and showed that the number of clumps and their masses are not excessive compared to observed galaxies at $z=2$. In addition to the chemical bimodality, the stars vertically scattered by the clumps built a geometric thick disk in excellent agreement with MW trends \citep{BeraldoeSilva2020}. Since only some of the early stars formed in clumps, both disks in the simulation formed simultaneously \citep[although the thick disk initially has a much higher star formation rate -- see Fig. 11 of][]{Clarke2019}, until clumps stopped appearing after about 4 Gyr, after which stars formed only in the thin disk. This scenario is supported by evidence for an old chemical thin disk population in the MW, as shown by \cite{BeraldoeSilva2021}.

In this paper, we track clumps in simulations of isolated MW-like galaxies to investigate their characteristics, including their SFRs, masses, lifetimes, locations, and chemical evolution over time. We consider two MW simulations that produced different levels of clumpiness, that both show an $\alpha$-abundance bimodality similar to that of the MW. Using both simulations, we find clumps of new stars in the first 4 Gyr and track their properties.

The layout of this paper is as follows.
In \autoref{sec:simulation}, we describe the simulations used in this study. In \autoref{sec:methods}, we present our clump detection and tracking algorithms. In \autoref{sec:results}, we summarize the properties of clumps in the two simulations. A discussion of the results is presented in \autoref{sec:discussion}, and, finally, our conclusions are presented in \autoref{sec:conclusion}.

\begin{figure*}[ht]
    \centering
    \includegraphics[width=0.8\textwidth]{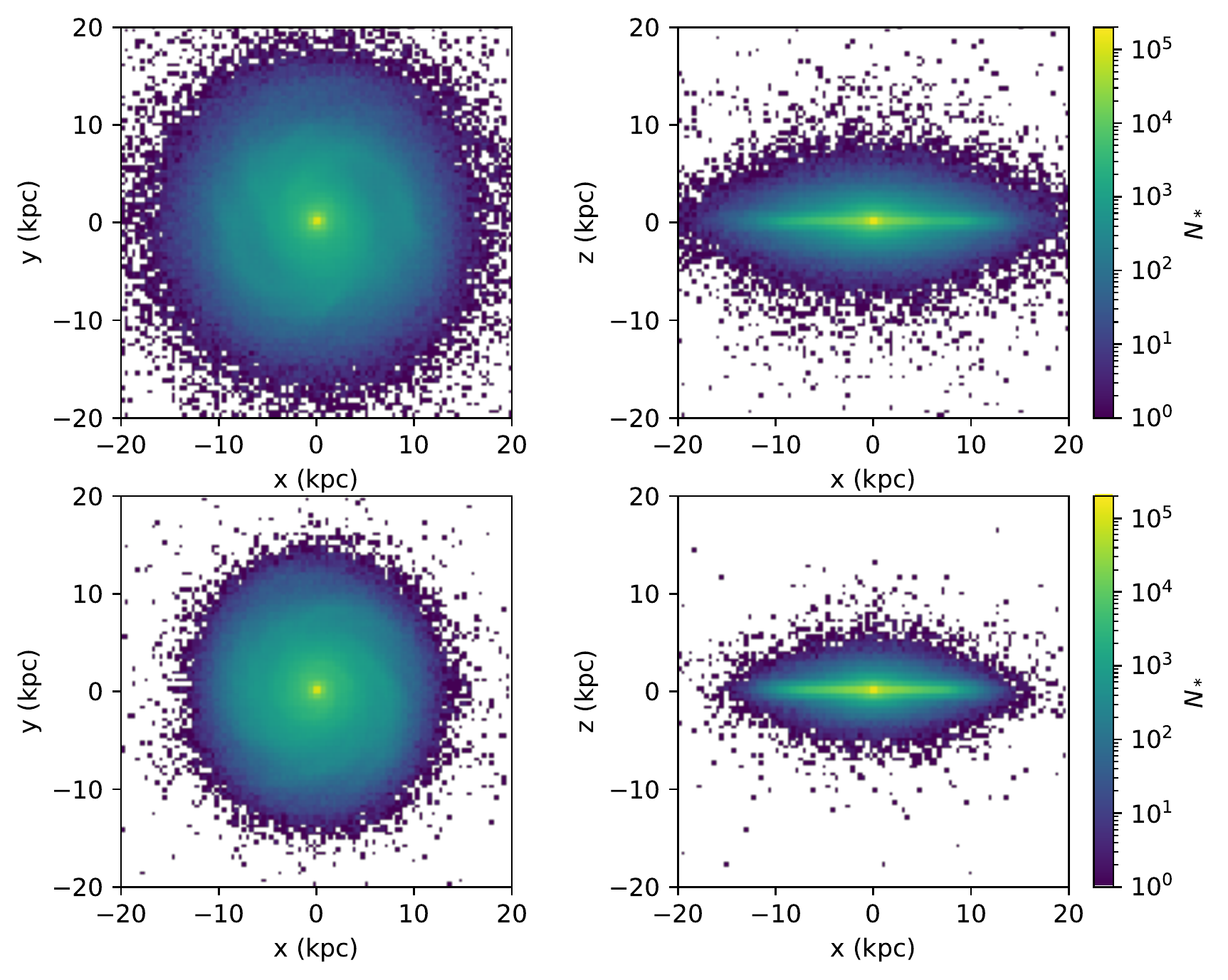}
    \caption{The face-on (left) and edge-on (right) views of the FB10 (10\% feedback) simulation (top) and the FB20 (20\% feedback) simulation (bottom) after 10 Gyr. The color shows the number of particles in each bin.}
    \label{fig:sims}
\end{figure*}

\begin{figure}
    \centering
    \includegraphics[width=0.95\columnwidth]{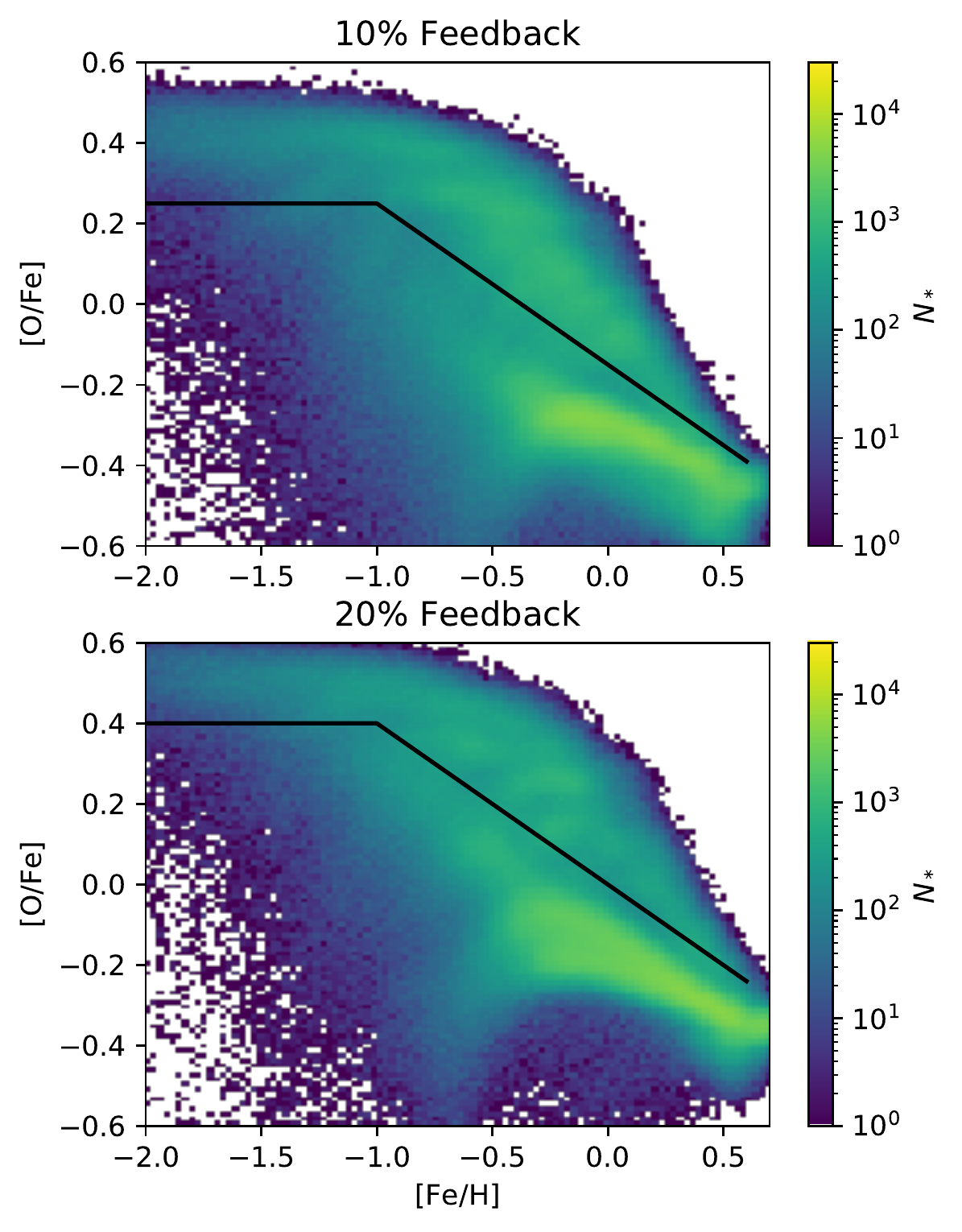}
    \caption{Chemistry of all star particles in the two simulations with the division between high and low-$\alpha$ regions marked by the black lines: ({\em top}) 10\% feedback (FB10), and ({\em bottom}) 20\% feedback (FB20). Color shows the number of particles in each bin.}
    \label{fig:alphadivision}
\end{figure}

\section{Simulations}
\label{sec:simulation}

The first simulation we use was previously described and investigated in detail by \citet{Clarke2019}, \citet{BeraldoeSilva2020} and \citet{Amarante2020}. The simulation starts with $10^6$ particles each of gas and dark matter, with the dark matter arranged in a spherical halo with a Navarro-Frenk-White \citep{Navarro1997} halo having a mass of $10^{12}$ \msun and virial radius of 200 kpc. The force softening of the model is 50 pc for gas and star particles, and 100 pc for the dark matter particles. The gas cools via metal-line cooling \citep{Shen2010} and stars form from a gas particle when it cools to 15,000 K and the density is greater than 1 cm$^{-3}$ in a converging flow. The simulation assumes a Miller-Scalo initial mass function (IMF). The mass resolution of the gas particles is $1.4 \times 10^5$ \msun and star particles form with a mass one third of this value. The galaxy is isolated with no accretion of subhalos. The number of Type Ia supernovae that occur over a certain time interval $t_b$ to $t_e$ is given by:
\begin{equation}
    \Delta N = \int_{m(t_e)}^{m(t_b)}\tilde{\Phi}(M_2) d M_2
\end{equation}
where $m(t)$ is the mass of stars that die at time $t$ and $\tilde{\Phi}(M_2)$ is given by
\begin{equation}
    \tilde{\Phi}(M_2) = \int_{M_{inf}}^{M_{sup}} \frac{24 f_B}{M_B}\mu^{\gamma} \Phi(M_B) d M_B/(M_B log(10)),
\end{equation}
where $M_{inf}=max(2M_2,3M_{\odot})$, $M_{sup}=M_2+8M_{\odot}$, $f_B$ is the normalization factor for the IMF, $\mu=M_2/M_B$, $\gamma=2$, and $\Phi$ is the IMF. In the first simulation (which we refer to hereafter as FB10), 10\% of the $10^{51}$ ergs of energy from supernovae is injected into the interstellar medium as thermal energy. The second simulation, which has not been previously presented, evolves the same initial conditions with 20\% of the energy from supernovae being injected into the interstellar medium and is hereafter termed model FB20. The models are evolved for 10 Gyr using the smooth particle hydrodynamics+$N$-body tree-code \textsc{gasoline} \citep{Wadsley2004,Wadsley2017}. \autoref{fig:sims} shows face-on and edge-on views for both models at $10$ Gyr. 

The simulation tracks [Fe/H] and [O/Fe] in the gas and star particles, which are based on the yield prescriptions of \citet{Thielemann1986},  \citet{Weidemann1987}, and \citet{Raiteri1996}. The yields from SNII are:
\begin{equation}
    M_{Fe} = 2.802 \times 10^{-4} M_*^{1.864}
\end{equation}
\begin{equation}
    M_O = 4.586 \times 10^{-4} M_*^{2.721}
\end{equation}
For SNIa, 0.13 \msun of O and 0.63 \msun of Fe are produced, regardless of the mass of the progenitor. These yield prescriptions result in [Fe/H] distributions that are a good match to those of the Milky Way \citep[see][]{Loebman2016}. For this work, we do not require that the values of the $\alpha$-abundances match those of the Milky Way in absolute terms, but only in relative terms with possible offsets, which these yield prescriptions provide.

In the FB10 model, we define high-$\alpha$ stars as those above a boundary and low-$\alpha$ stars as those below a boundary given by the relation:
\begin{equation}
\text{[O/Fe]}=
\begin{cases}
0.25 &\text{if [Fe/H]$<$-1.0} \\
-0.4\text{[Fe/H]}-0.15 &\text{otherwise}.
\end{cases}
\end{equation}
The FB20 model is shifted in chemical space relative to the FB10 simulation, and the division between high- and low-$\alpha$ for this model is: 
\begin{equation}
\text{[O/Fe]}=
\begin{cases}
0.4 &\text{if [Fe/H]$<$-1.0} \\
-0.4\text{[Fe/H]} &\text{otherwise}.
\end{cases}
\end{equation}
These divisions, on top of the resulting chemical maps, are shown as solid lines in \autoref{fig:alphadivision}.

\section{Methods}
\label{sec:methods}

We track the clumps in the simulations in order to investigate the evolution of their properties.
We define clumps as overdensities of new stars that last for at least 25 Myr.

Snapshots of the simulations are available every 100 Myr throughout the simulation. These snapshots include the position, velocity, and chemistry of each star particle at those times. The simulations also track the birth time and location of each star particle. This gives a higher time resolution for young stars than available in the snapshots. We use the information on where and when stars formed to find clumps.

\begin{figure}[h]
    \centering
    \includegraphics[width=0.95\columnwidth]{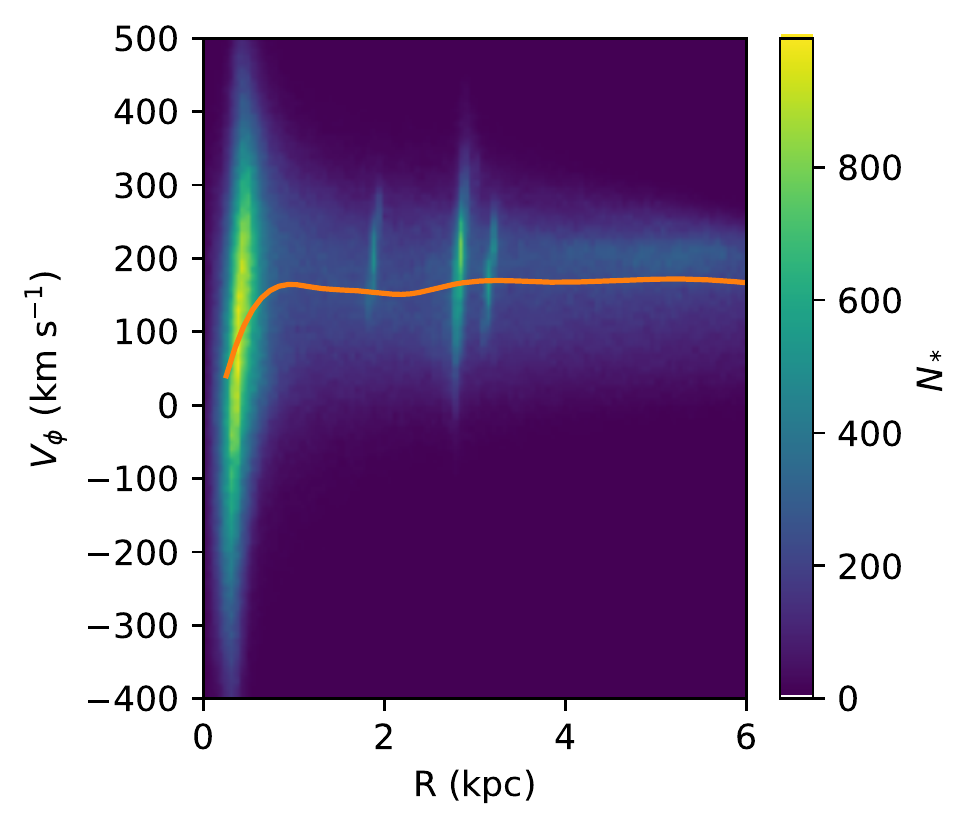}
    \caption{Rotation curve at 2.5 Gyr for model FB10. The nearly vertical stripes at radii between 2 and 4 kpc are clumps.}
    \label{fig:angular_velocity}
\end{figure}

\begin{figure}
    \centering
    \includegraphics[width=0.95\columnwidth]{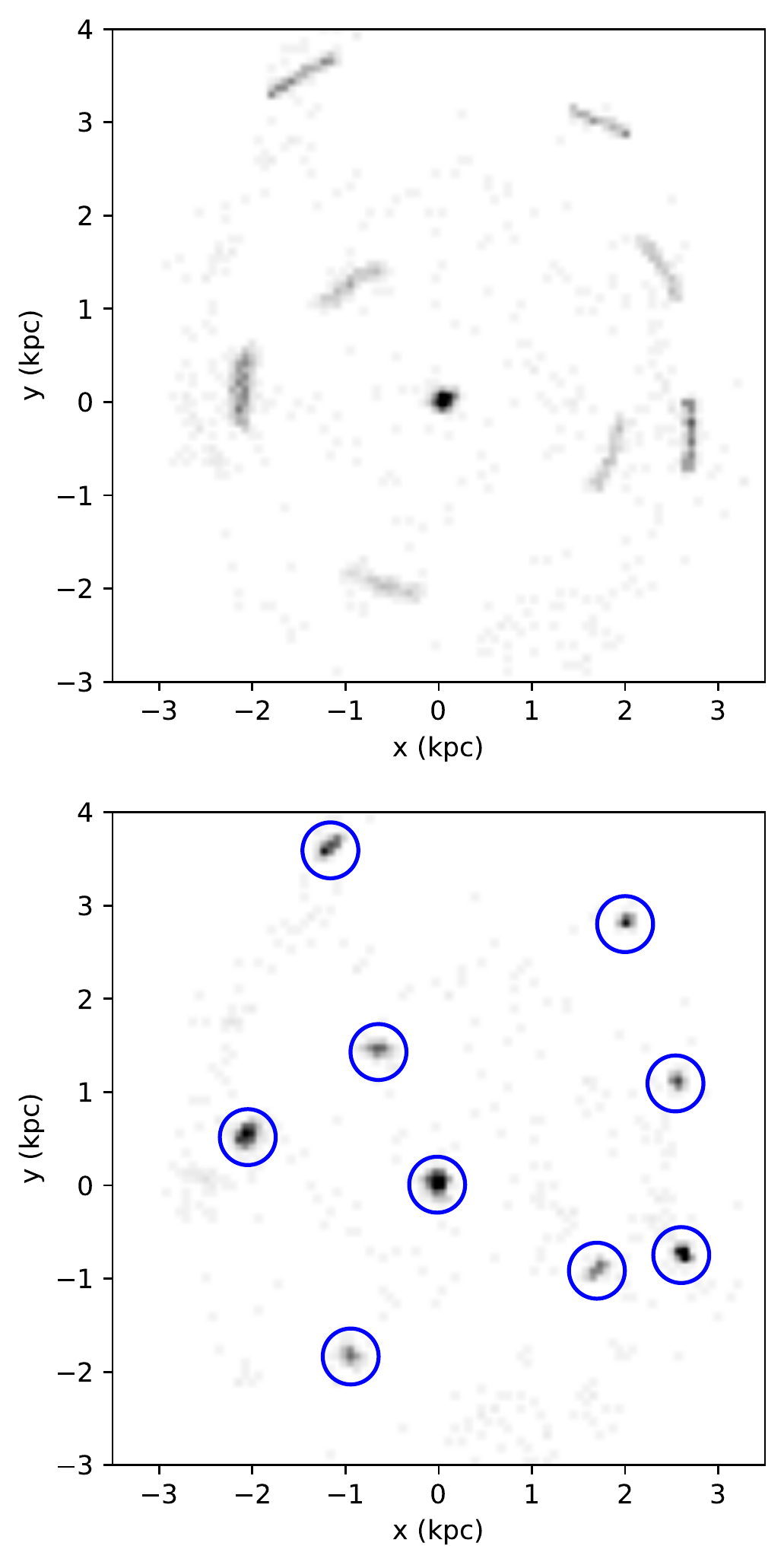}
    \caption{The stars that form within a 5 Myr time interval at $t=0.5$ Gyr, before ({\em top}) and after ({\em bottom}) being rotated back to where they would have been at the beginning of the time interval. The 0.3 kpc radius circles on the bottom plot show detected clumps.}
    \label{fig:correction}
\end{figure}

To track the clumps across time, we look at stars that formed over 5 Myr time intervals. During this time interval the clumps will have moved with the disk rotation, so stars that formed in the same clump at different times will have formed at different locations and make the clump appear to smear in position. In order to correct this, we remove the motion caused by the rotation of the galaxy. For each snapshot, we interpolate the angular velocity as a function of distance from the axis of rotation using the median angular velocities of the stars with bins of width 0.5 kpc. \autoref{fig:angular_velocity} shows this curve at 2.5 Gyr for the FB10 simulation. We use this rotation curve to move stars that formed within a 5 Myr time interval back to where they would have been at the beginning of that interval in order to correct for the rotation of the galaxy over that time, as shown in \autoref{fig:correction}. The values for the angular velocity are median values for their bins; since there is a high velocity dispersion, especially near the center and within clumps, we do not use time intervals longer than 5 Myr.

To detect clumps, we perform the following five steps at every 1 Myr between 0 and 4 Gyr:

\begin{itemize}
    \item We find all the stars that formed within the 5 Myr time interval;
    \item We then move the stars back to where they would have been at the beginning of the time interval, using the angular velocity curve;
    \item We find high density peaks in the face-on density plot of these stars;
    \item We combine peaks within 0.3 kpc of each other and set the center of the combined peaks at the mean $x$ and $y$ location of nearby stars;
    \item We find stars that formed within 0.3 kpc of the clump center.
\end{itemize}

To find areas with high star formation rate density, we produce an $x$--$y$ (face-on) density map with square bins with side length $0.15$ kpc. Then we use the function \texttt{find\_peaks()} from \texttt{photutils} \citep{Bradley2019} with a threshold of $2.5\sigma$ above the mean of non-zero bins to detect peaks. Since a single clump could include multiple bins above the threshold, we combine peaks within 0.3 kpc of each other.

To track clumps between time intervals, we match each clump to those found within 0.3 kpc, after removing galactic rotation, in the five most recent time steps. Clumps that appear for less than 25 Myr are considered noise and removed.

\begin{figure}[ht]
    \centering
    \includegraphics[width=0.95\columnwidth]{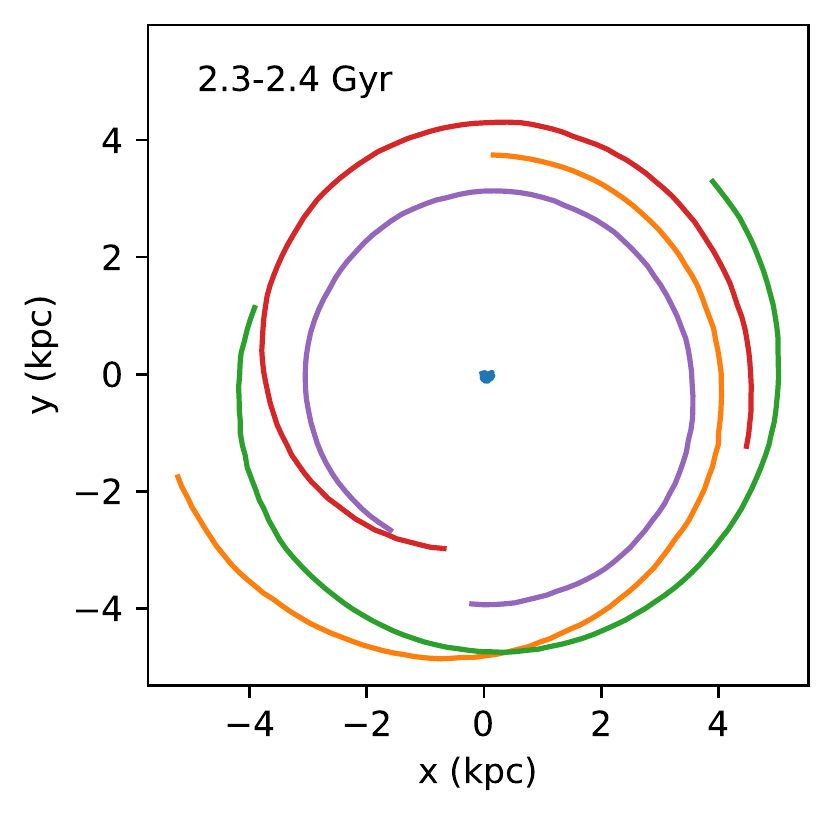}
    \includegraphics[width=0.95\columnwidth]{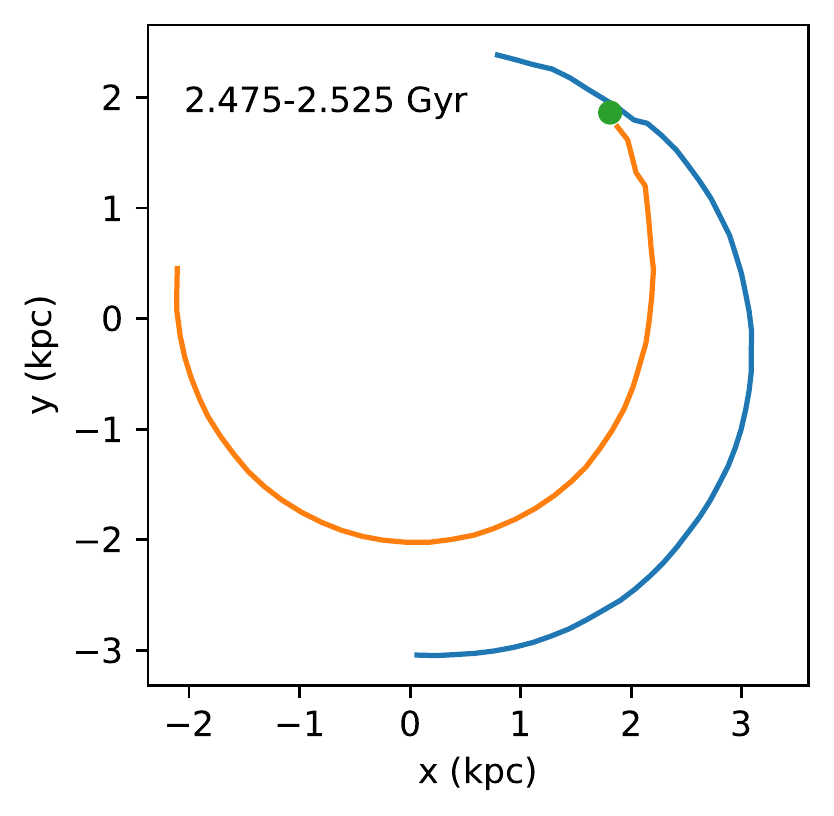}
    \caption{({\em Top}) The paths of several clumps in the FB10 simulation over a 100 Myr period. The clumps move counter-clockwise and many of them eventually spiral into the center. ({\em Bottom}) The paths of two clumps that merge. The dot shows where they merge.}
    \label{fig:clumppaths}
\end{figure}

If a clump matches with two or more clumps from previous iterations, then they are considered to have merged. Mergers between clumps are fairly common, especially in the first 1.5 Gyr. Usually two or three clumps merge at once, and many of the clumps eventually merge into the galactic center, populating the bulge -- see \citet{Debattista2023}. \autoref{fig:clumppaths} shows an example of paths of several clumps over 100 million years, as well as the paths of two clumps that merge.

To track the chemistry of the clumps over time, and determine whether they crossed into the high-$\alpha$ region, we measure the mean \feh~ and \ofe~ abundances of all star particles that formed in those clumps over a 5 Myr time interval.

\section{Results}
\label{sec:results}

\begin{figure*}
    \centering
    \includegraphics[width=0.85\textwidth]{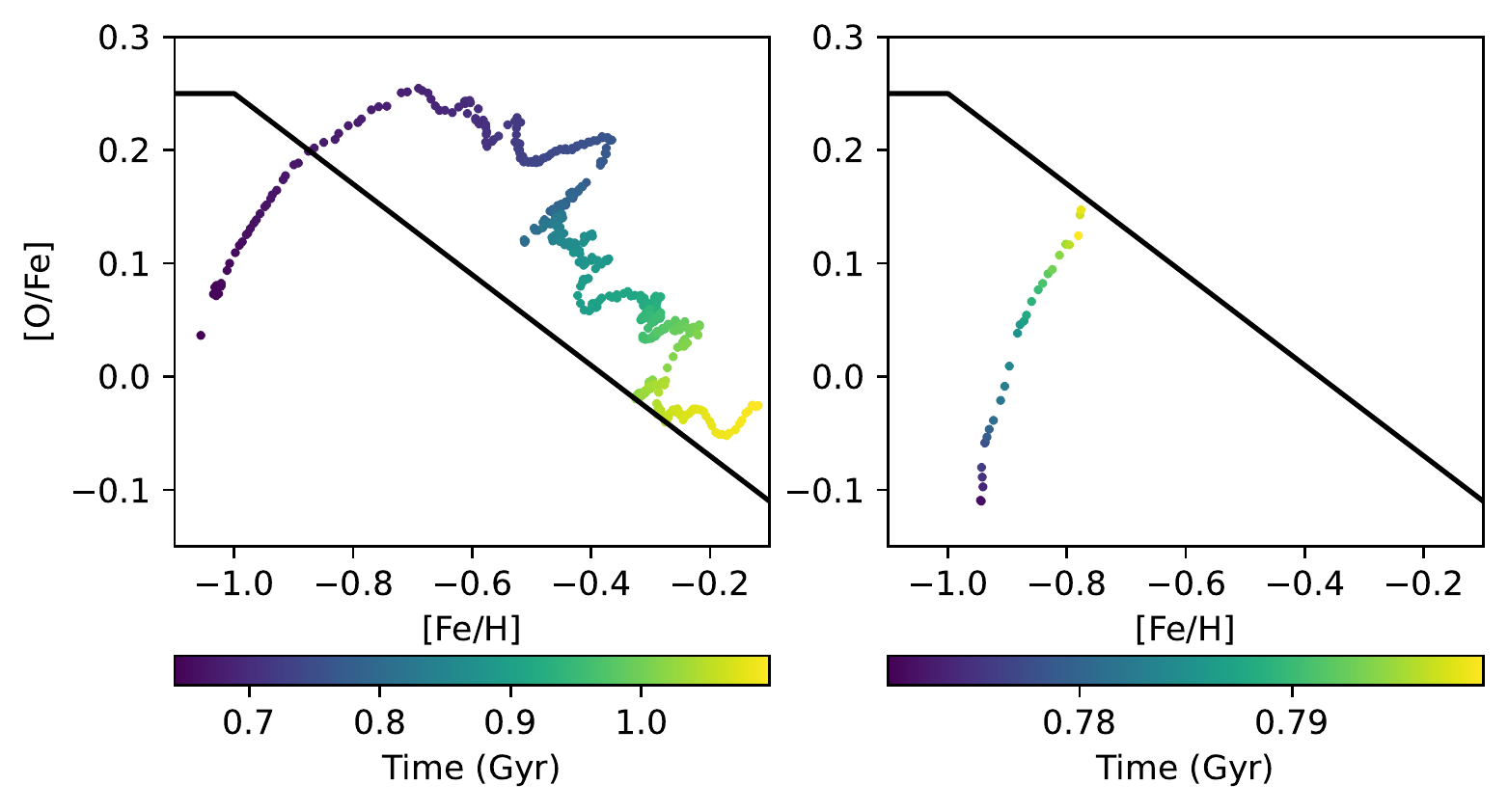}
    \caption{The paths of two typical clumps in FB10 in \ofe-\feh~ space. The abundance values for each clump are the mass-weighted mean of stars that formed over 5 Myr. One of them (left) reaches the high-$\alpha$ track while the other one (right) does not. The division between high- and low-$\alpha$ is also shown. The color indicates the time.
    \label{fig:clumpchem}}
\end{figure*}

\begin{figure*}
    \centering
    \includegraphics[width=0.85\textwidth]{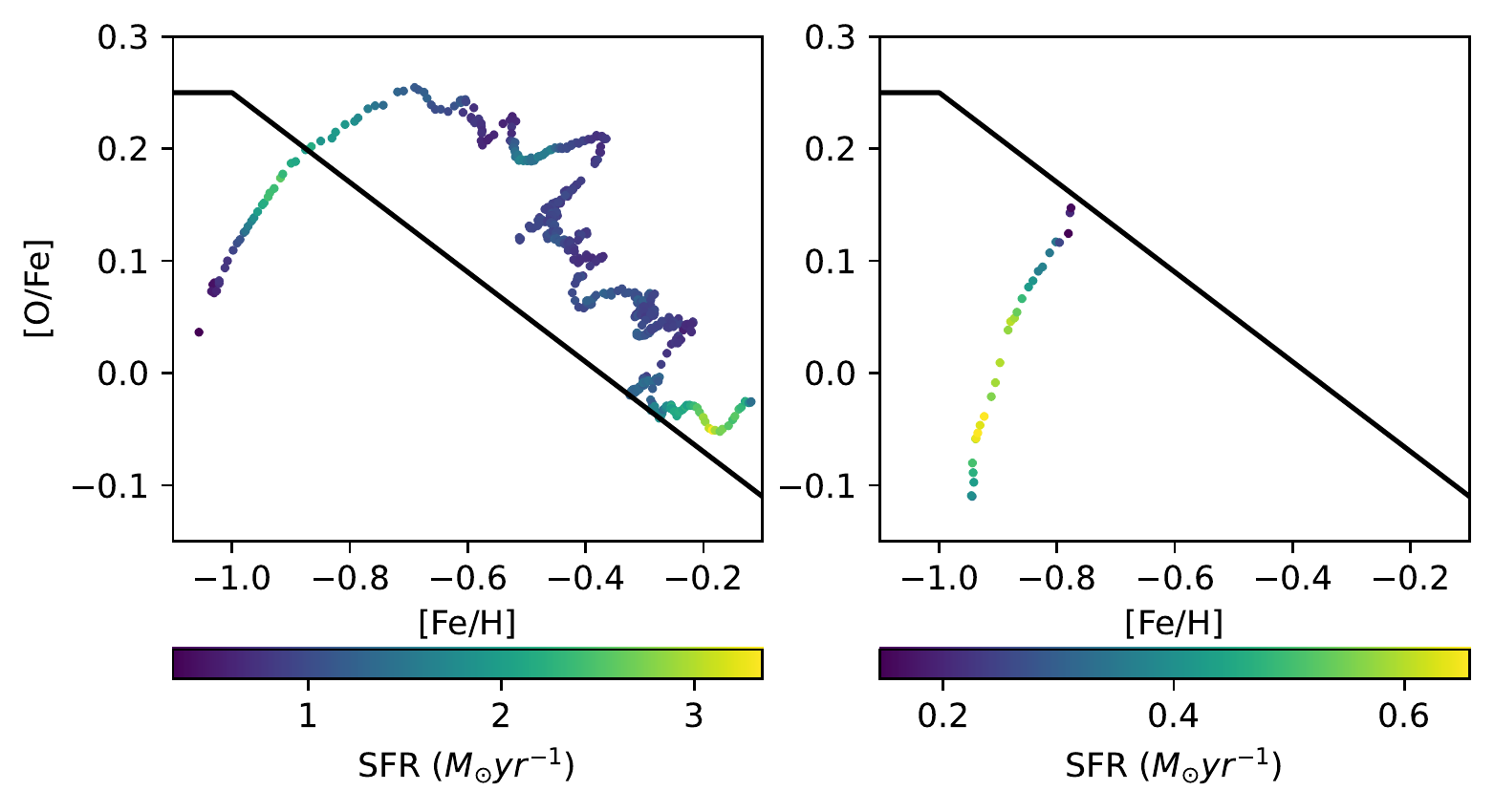}
    \caption{Same as \autoref{fig:clumpchem} but colored by SFR.}
    \label{fig:clumpchemsfr}
\end{figure*}

We detect and track exactly 100 clumps in the FB10 simulation during the first 4 Gyr, after which we find no further clumps. Of these, 68 reached the high-$\alpha$ track and 36 have a lifetime of at least 100 Myr. As the example shown in \autoref{fig:clumpchem} and \autoref{fig:clumpchemsfr} illustrates, most  of the clumps start on the low-$\alpha$ side (though some start with high $\alpha$ abundance) and quickly increase in \ofe~ at nearly constant \feh~ at first. As shown in \autoref{fig:clumpchemsfr}, the SFR is high during this phase, and decreases after the clump reaches high-$\alpha$. While some of the clumps eventually cross the boundary into the high-$\alpha$ region (high-$\alpha$ clumps), others do not (low-$\alpha$ clumps). Most of the high-$\alpha$ clumps then decrease in \ofe~ as their metallicity increases, i.e. similar to a ``standard'' chemical evolution track once SNIa feedback has turned on. As shown in the figure, the high-$\alpha$ clump survives for about 400~Myr while the low-$\alpha$ clump, shown in the right panels of \autoref{fig:clumpchem} and \autoref{fig:clumpchemsfr} only survives for about 30~Myr which does not give it enough time to cross over to the high-$\alpha$ side.

In the FB20 simulation, we find 92 clumps. Of those, 59 reach the high-$\alpha$ track. Many of the clumps that reach the high-$\alpha$ track either fully dissolve or reach the galactic center before their metallicity increases substantially.

\begin{figure}
    \centering
    \includegraphics[width=0.95\columnwidth]{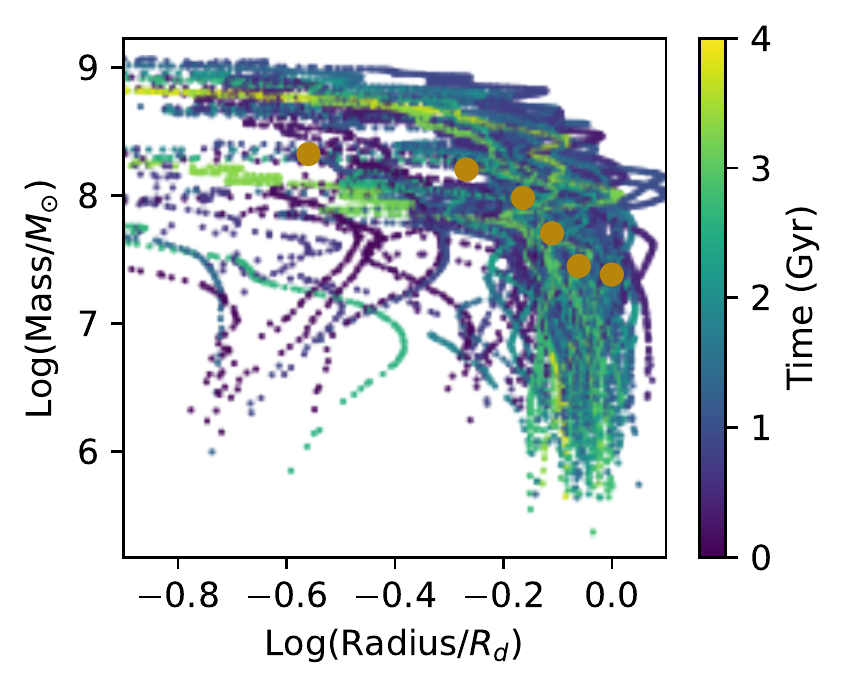}
    \includegraphics[width=0.95\columnwidth]{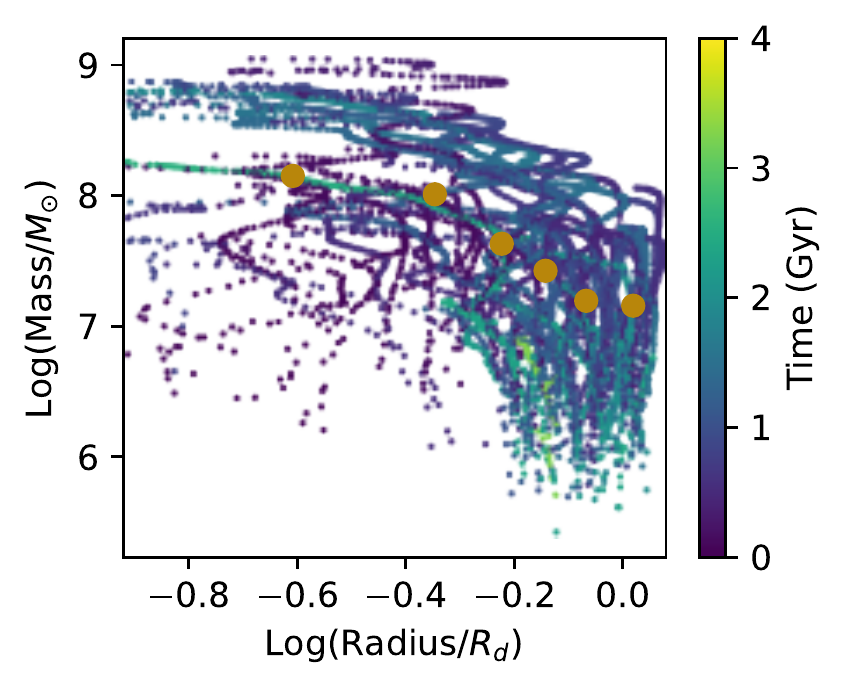}
    \caption{Evolution of the stellar mass vs. radius for each clump, colored by time, with larger dots showing the mean for radial bins containing equal numbers of points, ({\em top}) in the FB10 simulation, and ({\em bottom}) in the FB20 simulation. Clumps increase in mass as they fall toward the center, so the ones at smaller radii tend to be more massive by a factor of about 10 between the inner and outer galaxy.}
    \label{fig:massvr}
\end{figure}

\begin{figure}
    \centering
    \includegraphics[width=0.95\columnwidth]{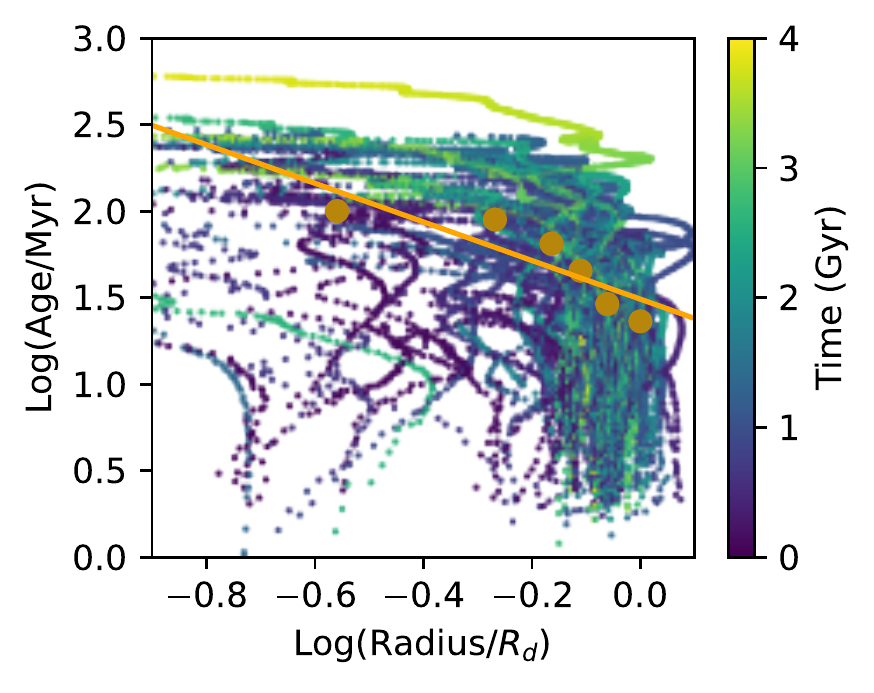}
    \includegraphics[width=0.95\columnwidth]{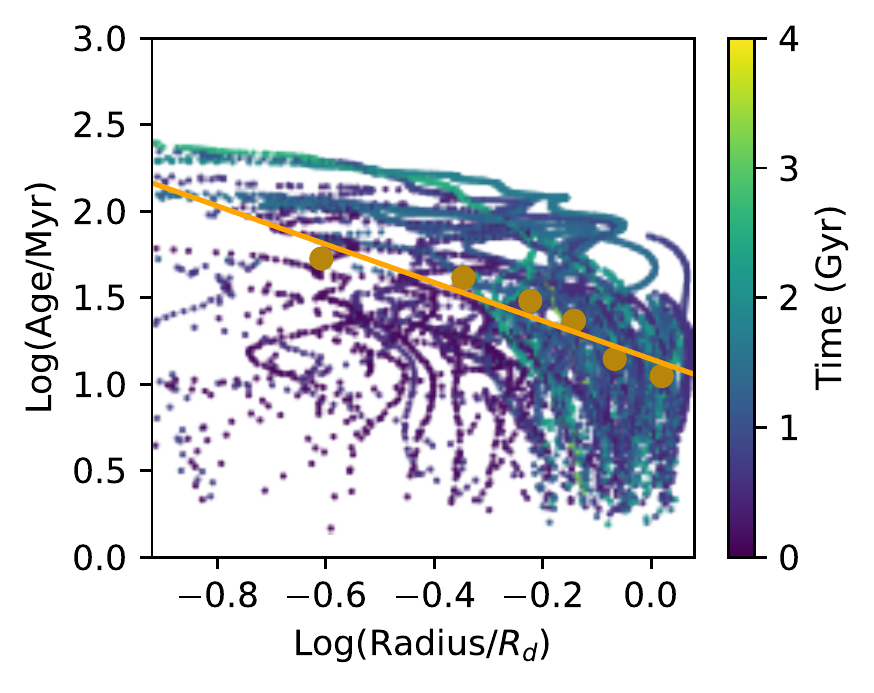}
    \caption{Evolution of mean stellar age vs. radius for each clump, colored by time, with larger dots showing the mean for radial bins containing equal numbers of points, ({\em top}) in the FB10 simulation, and ({\em bottom}) in the FB20 simulation. The trend lines show the gradients. Clumps closer to the galactic center tend to be older than clumps at larger radii. The gradient for FB10 is -1.12 and the gradient for the FB20 simulation is -1.11.}
    \label{fig:agevr}
\end{figure}

Most of the high-$\alpha$ clumps in both simulations reach the center, though there are quite a few that dissolve instead. Of the 59 clumps in the FB20 simulation that reach high-$\alpha$, 32 eventually merge into the center (54\%), whereas in the FB10 simulation, 42 of the 68 high-$\alpha$ clumps merge into the center (62\%). Since the clumps on average move toward the center and increase in mass as they go, the mean mass and mean age of clumps is higher for a lower galactic radius, as shown in \autoref{fig:massvr} and \autoref{fig:agevr}, respectively. In these figures, the radius is normalized by $R_d$, which is the radius that contains 85\% of the mass of stars less than 100 Myr old and gas with a temperature below $1.5\times 10^4$ K, as defined by \citet{Mandelker2017}.

\begin{figure}
    \centering
    \includegraphics[width=0.95\columnwidth]{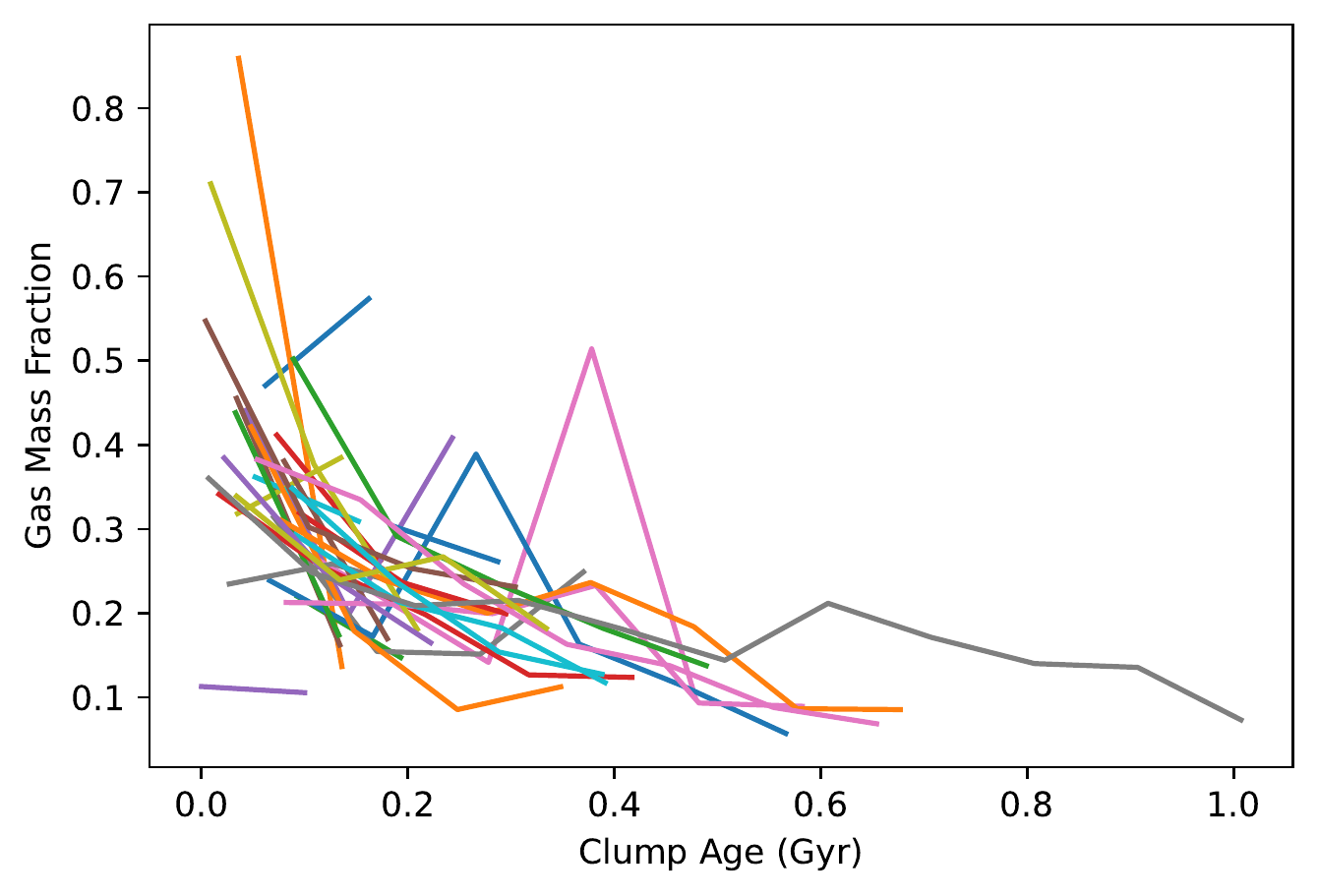}
    \includegraphics[width=0.95\columnwidth]{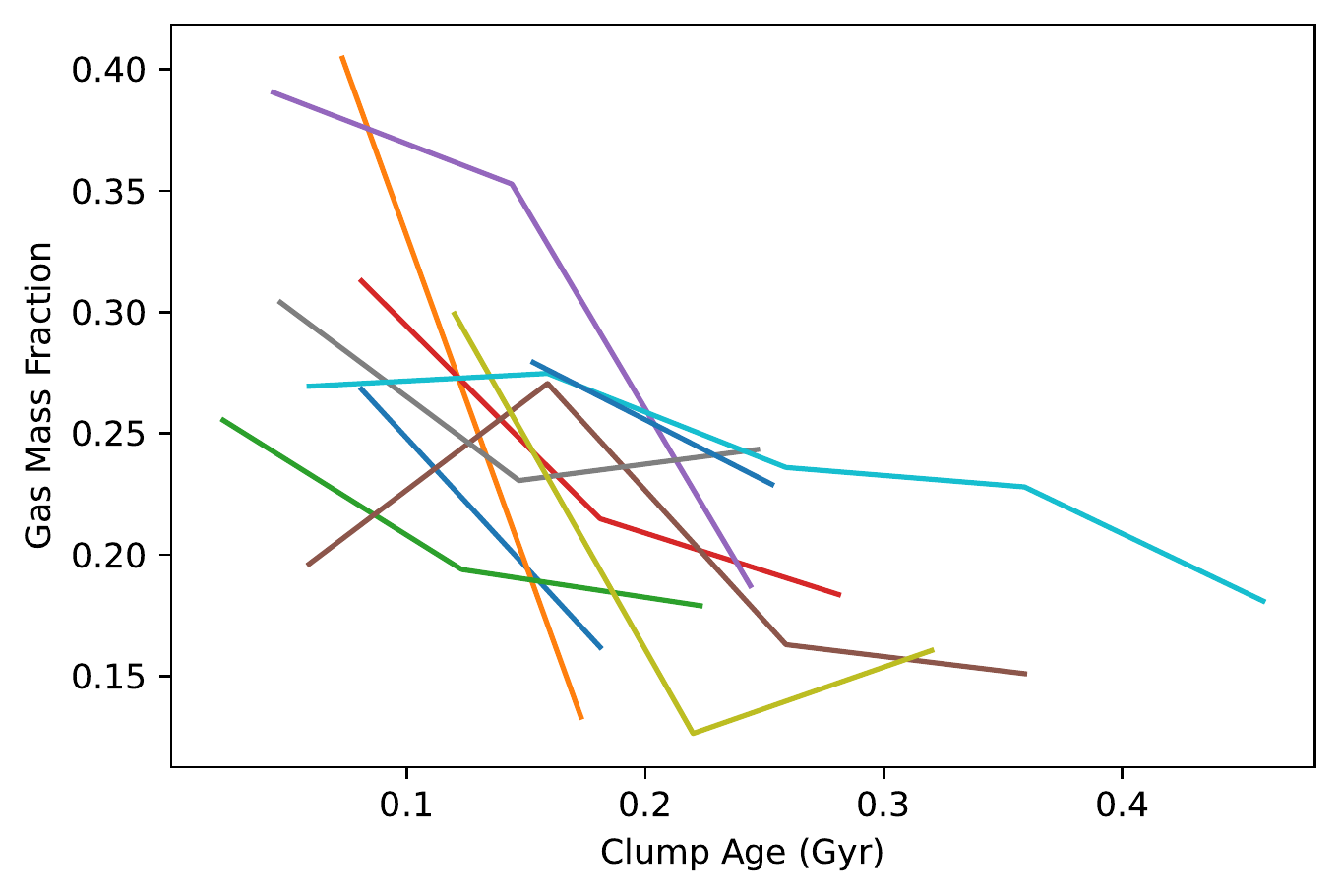}
    \caption{Evolution of gas mass fraction vs. time since the clump was first detected for clumps that are present for at least two snapshots, ({\em top}) in the FB10 simulation, and ({\em bottom}) in the FB20 simulation. Most of the clumps have a gas mass fraction of less than 0.5 at the time of the first snapshot after detection. This may be due to the fact that we detect clumps once they have started forming stars rather than first detecting them as an instability in the gas.}
    \label{fig:gasfrac}
\end{figure}

In both simulations, the gas mass fraction within clumps starts out high and decreases on average over a clump's lifetime. After about 200 Myr, it drops to around 0.3 or less for most clumps. The gas fractions for clumps in both simulations are shown in \autoref{fig:gasfrac}.

\subsection{Model FB10} 

\begin{deluxetable*}{cccccc}
    \tablecolumns{6}
    \tablecaption{Mean Clump Properties in Both Simulations. The SFR is more than 3.3 times higher, the log of lifetime and log of mass are both more than 1.1 times higher on average for the high-$\alpha$ clumps than for the low-$\alpha$ clumps in both simulations. High-$\alpha$ clumps on average form 1.4 kpc closer to the Galactic center in FB10 and 1.5 kpc closer to the center in FB20.} \label{table:1}
    \tablehead{\colhead{Simulation} & \colhead{$\alpha$ Abundance} & \colhead{SFR} & \colhead{Log(Lifetime/1 Myr)} & \colhead{Log(Mass/\msune)} & \colhead{Initial Radius} \\ 
    \colhead{} & \colhead{} & \colhead{(\msun yr$^{-1}$)} & \colhead{} & \colhead{} & \colhead{(kpc)} }
    \startdata
    FB10 & high & $1.20\pm0.73$ & $2.13\pm0.41$ & $8.13\pm0.46$ & $3.19\pm1.47$ \\
    & low & $0.36\pm0.23$ & $1.67\pm0.14$ & $7.15\pm0.29$ & $4.6\pm1.49$ \\
    \hline
    FB20 & high & $1.39\pm0.71$ & $1.88\pm0.32$ & $7.97\pm0.40$ & $2.44\pm1.31$ \\
    & low & $0.39\pm0.2$ & $1.66\pm0.15$ & $7.19\pm0.26$ & $3.96\pm1.28$ \\
    \enddata
\end{deluxetable*}

\begin{figure}
    \centering
    \includegraphics[width=0.95\columnwidth]{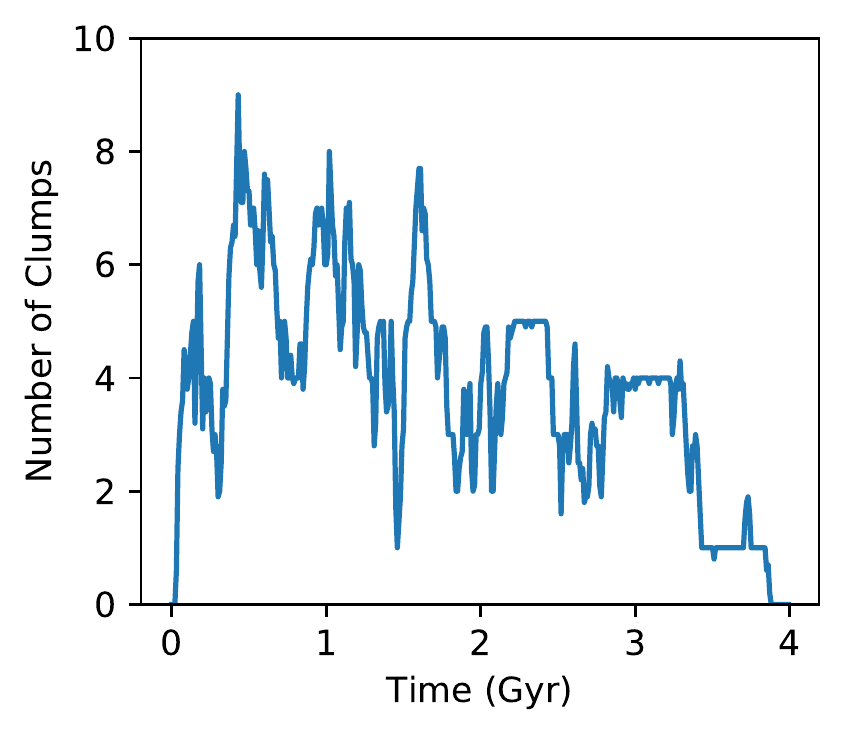}
    \includegraphics[width=0.95\columnwidth]{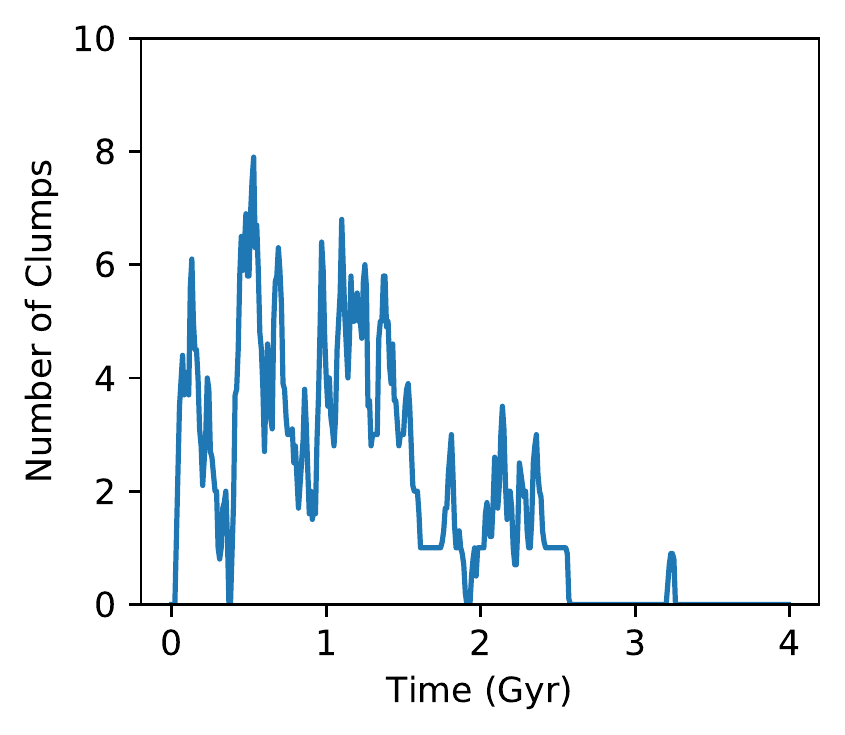}
    \caption{The number of clumps detected at each time (not counting the center), averaged over 10 Myr, ({\em top}) in the FB10 simulation, and ({\em bottom}) in the FB20 simulation. The number of clumps that appear at the same time is largest early in the run and decreases over time. Clumps stop forming in FB20 after $\sim 2.5$ Gyr and in FB10 after $\sim 4$ Gyr.}
    \label{fig:nclumps}
\end{figure}

\begin{figure*}[t]
    \centering
    \includegraphics[width=0.45\textwidth]{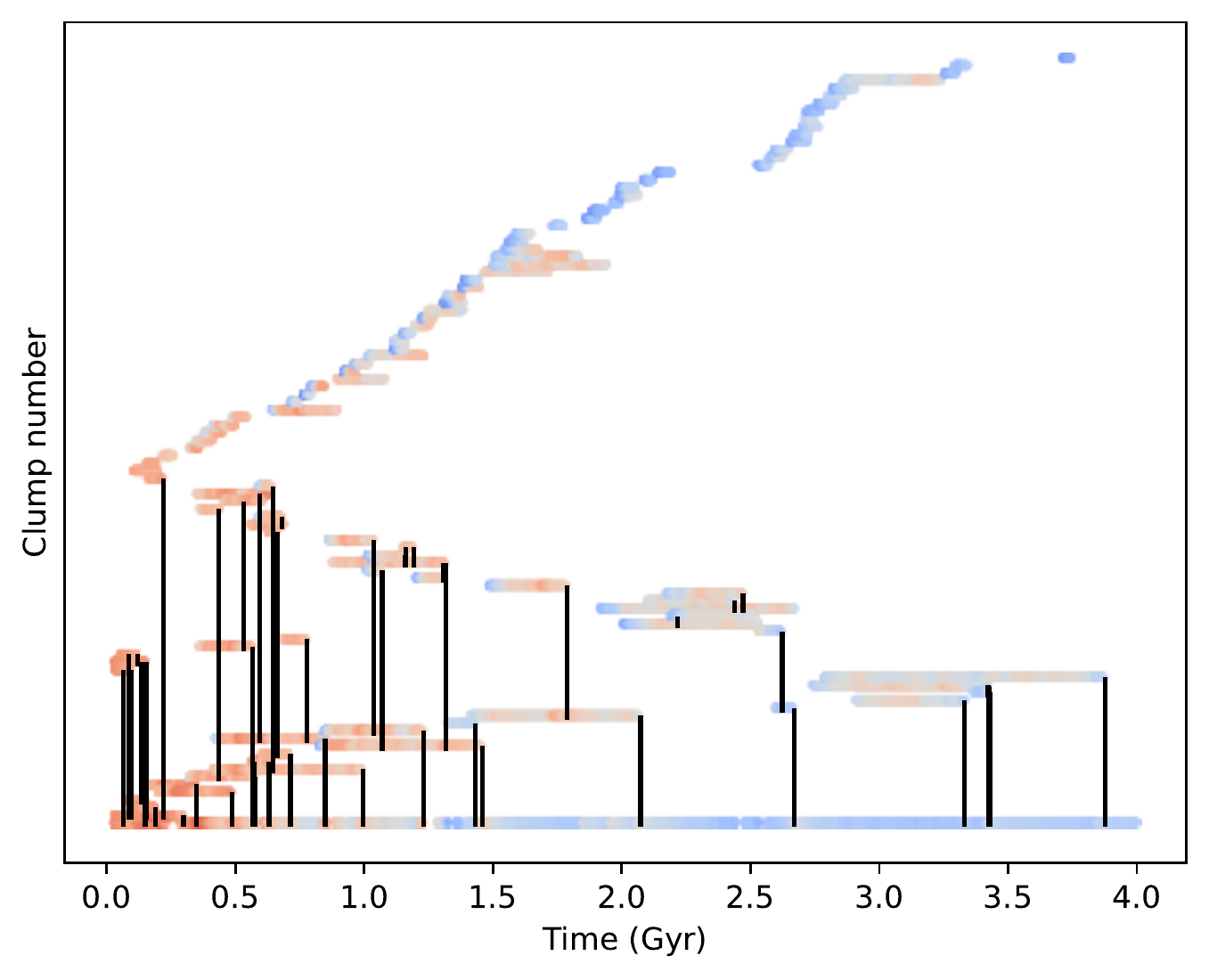}
    \includegraphics[width=0.51\textwidth]{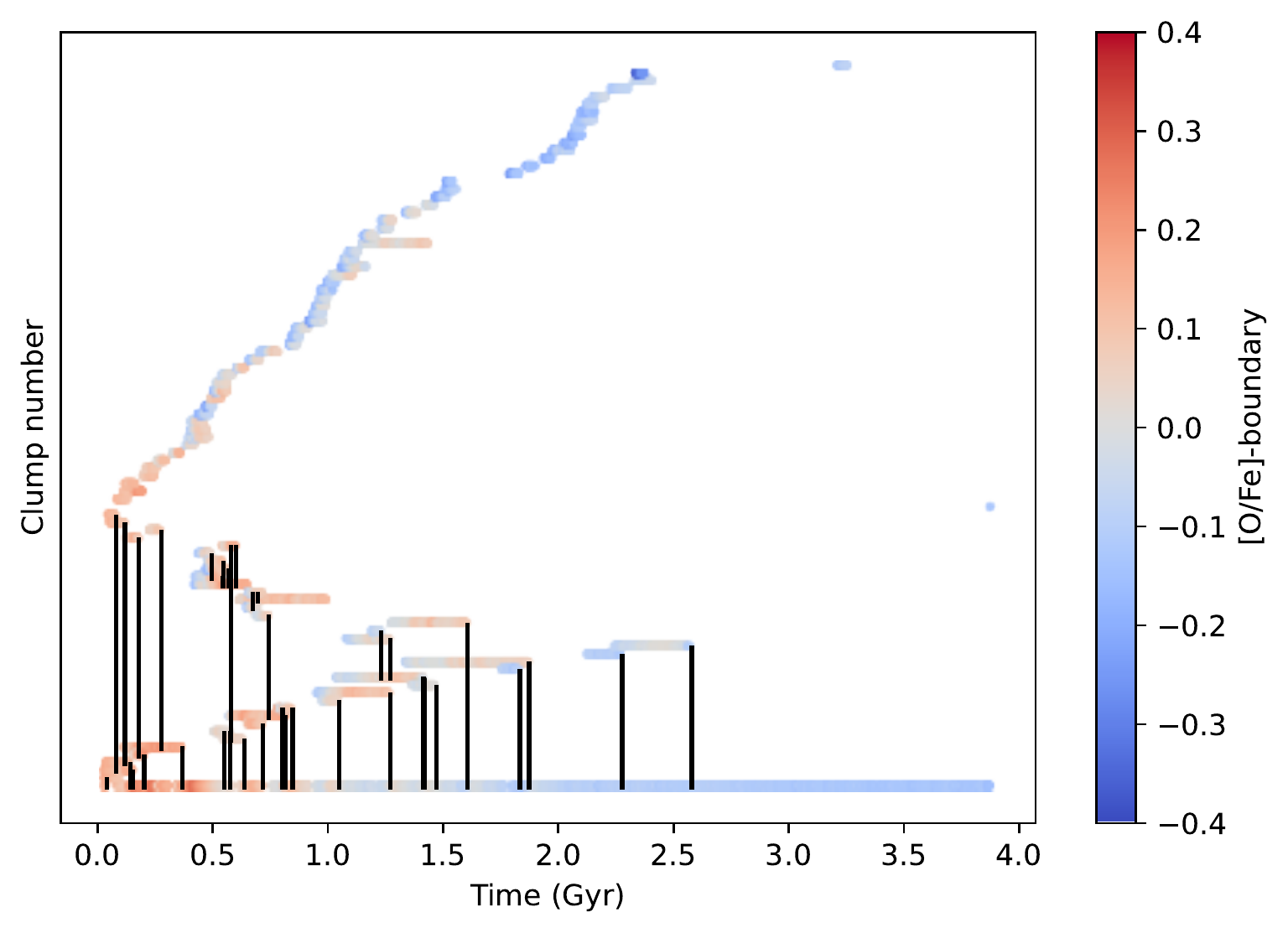}
    \caption{Merger trees for the FB10 model ({\em left}) and the FB20 model ({\em right}) showing high-$\alpha$ clumps in red and low-$\alpha$ clumps in blue. Each horizontal line corresponds to an individual clump. Vertical black lines denote mergers. The clump at the bottom is the galactic center. Color is the distance from the high-$\alpha$ cutoff line for the clump's metallicity.}
    \label{fig:mergertree}
\end{figure*}

\autoref{fig:nclumps} shows the number of clumps detected at each time for the FB10 (FB20) simulation at the top (bottom) panel. In the FB10 simulation, clumps appear during the first $\sim 4$ Gyr. For the first 1.5 Gyr, many clumps appear at nearly the same time, and merge into the center fairly regularly. Between 1.5 and 4 Gyr, there are still several clumps that appear at the same time, though the number starts to decline, as shown in \autoref{fig:nclumps}. Some clumps dissolve, while others merge with other clumps or fall into the center. This can be seen in \autoref{fig:mergertree} which shows the mergers between clumps and with the center. Each horizontal line on the plot represents one clump, and the black vertical lines represent mergers between clumps. After 4 Gyr, all clumps have either dissolved or moved to the center and no more clumps appear.

Many of the high-$\alpha$ clumps eventually spiral into the center, while continuously shedding star particles that spread throughout the galaxy. This is shown in \autoref{fig:mergertree}, where most of the high-$\alpha$ clumps are connected to the center through mergers, while most of the low-$\alpha$ clumps do not merge with any other clumps.

\begin{figure*}[t]
    \centering
    \includegraphics[width=0.9\textwidth]{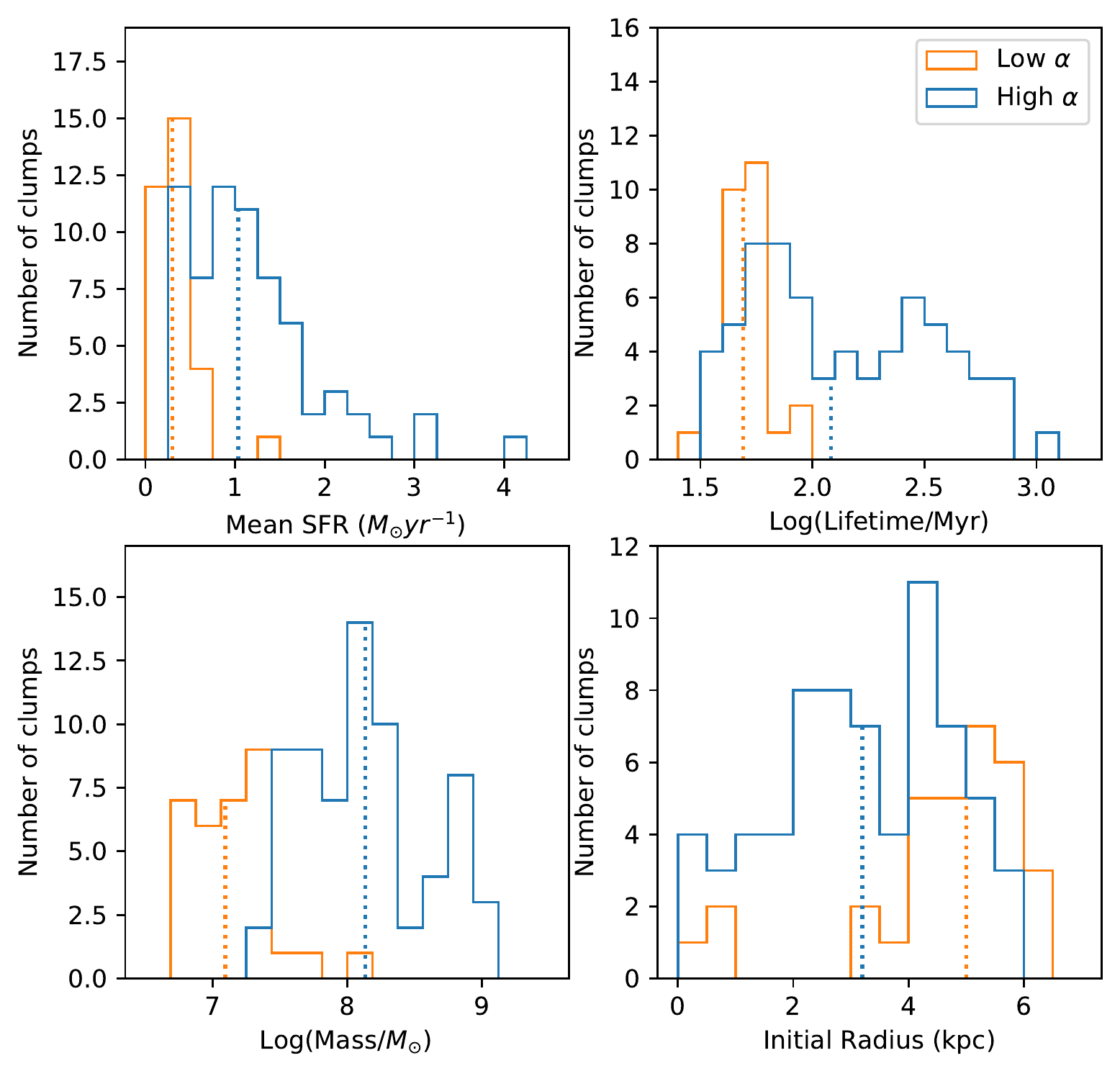}
    \caption{Histograms of properties of the high- and low-$\alpha$ clumps in the FB10 simulation. ({\em Top left}) Mean SFR in \msun yr$^{-1}$, ({\em Top right}) log of lifetime in Myr, ({\em Bottom left}) log of mass in \msune, and  ({\em Bottom right}) initial radius in kpc. The dotted lines show the median values for each group.}
    \label{fig:hists}
\end{figure*}

We compare the properties of clumps that reach the high-$\alpha$ track to those that do not. \autoref{fig:hists} shows the differences in mean SFR, lifetime, mass, and birth radius between the high-$\alpha$ and low-$\alpha$ clumps. Clumps that reach high-$\alpha$ tend to have a higher mean SFR, a greater mass, last longer, and form closer to the galactic center than those that do not.

\begin{figure*}[t]
    \centering
    \includegraphics[width=0.9\textwidth]{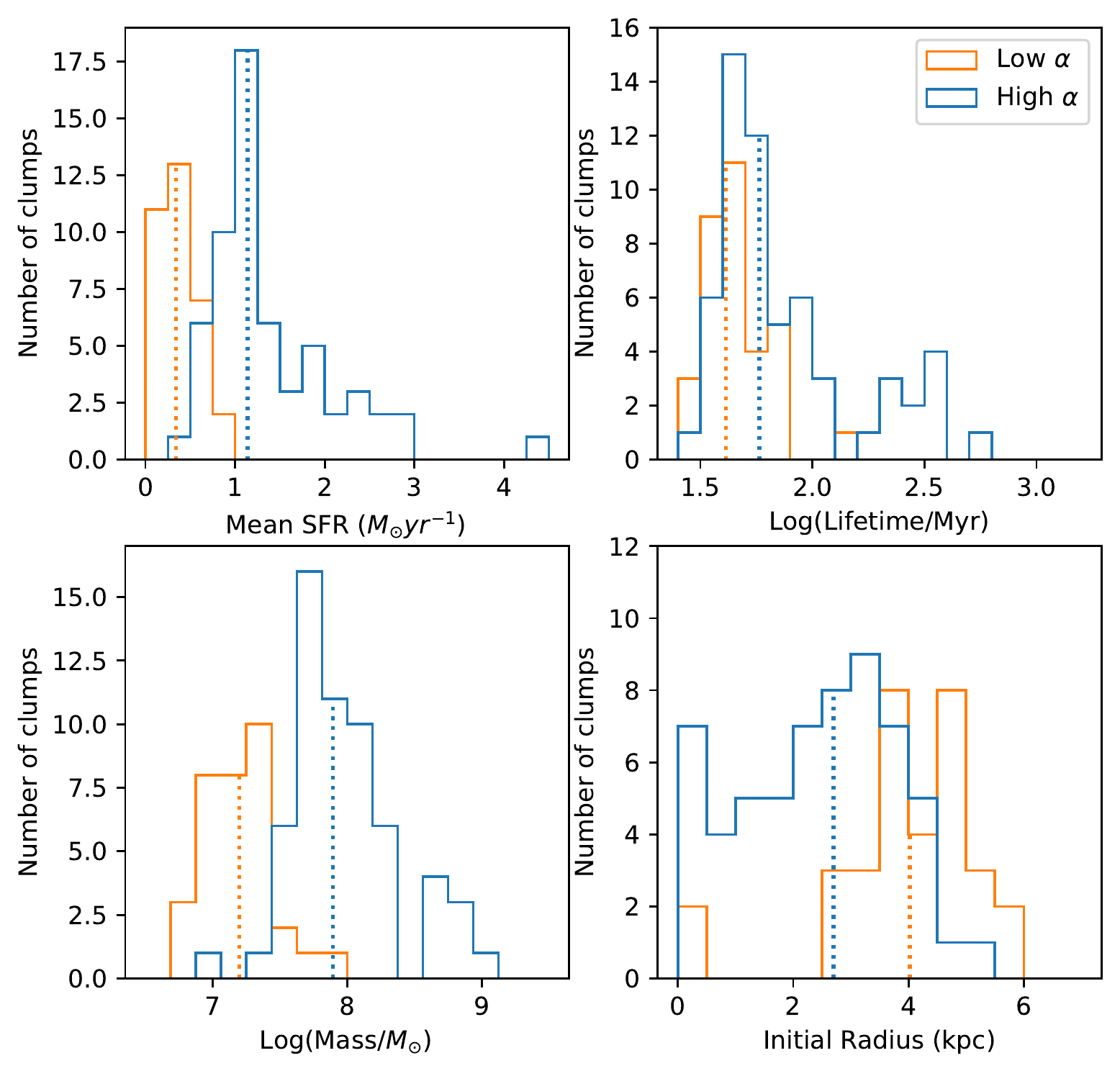}
    \caption{Same as \autoref{fig:hists} but in the FB20 simulation.}
    \label{fig:hists20}
\end{figure*}

\subsection{Model FB20}

The bottom panel of \autoref{fig:nclumps} shows that, in the FB20 simulation, clumps only appear for the first $\sim 2.5$ Gyr. On average, the clumps in FB20 have a shorter lifetime than the ones in FB10. \autoref{fig:hists20} shows that, as in the FB10 simulation, the high-$\alpha$ clumps on average have a higher mean SFR, longer lifetime, greater mass, and lower birth radius than the low-$\alpha$ clumps. \autoref{table:1} lists the mean and standard deviations of properties of high- and low-$\alpha$ clumps in both simulations.

Fewer clumps are present at any one time in the FB20 simulation than in the FB10 simulation, as seen in \autoref{fig:nclumps}. For the first $\sim 1.6$ Gyr, the FB20 simulation has a mean of 3.76 clumps and a maximum of 8 clumps at once. The FB10 simulation by comparison has a mean of 5 clumps and a maximum of 9 clumps appearing at once over the same time interval. After 1.6 Gyr, the number of clumps in the FB20 simulation drops to one, and it stays less than 4 until 2.58 Gyr, when clump formation ceases. One additional clump appears at $\sim 3.2$ Gyr, but it only lasts a few tens of Myr and does not reach high $\alpha$ or merge into the center.

\section{Discussion} \label{sec:discussion}

We have tracked clumps in two MW-like simulations where the only difference in parameters is the amount of feedback from supernovae and divided them into two groups: those that reach the high-$\alpha$ track, and those that do not. The clumps that last at least 100 Myr tend to reach the high-$\alpha$ track and eventually spiral into the galactic center.  In contrast, many of the clumps that do not last that long stay on the low-$\alpha$ track and dissolve rather than reaching the center. Clump formation continues later and clumps last longer in the 10\% feedback FB10 model than in the 20\% feedback FB20 model.

The fact that clumps in the FB20 simulation do not tend to last as long as those in the FB10 simulation implies that higher feedback disrupts clumps and causes them to dissolve sooner, as has been found in previous studies \citep{Hopkins2012,Oklopcic2017,Fensch2021}. \citet{Fensch2021} studied the effect of gas fraction and feedback on forming giant clumps in simulations. They found that clumps in simulations with higher feedback did not last as long as clumps in simulations with lower feedback, though a higher gas fraction had more effect on producing longer clump lifespans than a lower feedback.

The time over which the clumps appear is also different for the two simulations, as seen in \autoref{fig:nclumps}. In the FB10 simulation, clumps appear for the first $\sim 4$ Gyr of the galaxy's lifetime, which is $\sim 1.5$ Gyr longer than clump formation occurs in FB20. As a result, the thick disk stars in the FB20 simulation would be older on average than the thick disk stars in the FB10 simulation.

A common characteristic of the clumps in both simulations is that most of them eventually spiral into the galactic center. As they do so, they populate it with many high-$\alpha$ stars. In the FB10 simulation, 30.4\% of the high $\alpha$ stars end up within 0.5 kpc from the center and 44.3\% of the stars in the center have high $\alpha$ abundances. In the FB20 simulation, 39.7\% of the high-$\alpha$ stars end up within 0.5 kpc from the center and 42.8\% of the stars in the center have high $\alpha$ abundances. Other simulations have also shown clumps migrating toward the center to form the bulge \citep{Immeli2004,Dekel2009}. \cite{Debattista2023} studies the effect of star formation in clumps on the bulge of model FB10, finding that it results in a chemical distribution similar to the bulge of the MW.

Because clumps migrate to the center, the average mass and age of clumps is higher toward the center for both simulations, as shown in \autoref{fig:massvr} and \autoref{fig:agevr}. Observations have found that older clumps tend to be closer to the center of high-redshift galaxies than younger clumps \citep{ForsterSchreiber2011,Guo2012,Adamo2013,Guo2018,Lenkic2021}. \citet{Lenkic2021} find a gradient of -0.63 in the relation between log age and log radius, while \citet{Guo2018} finds gradients ranging from -2.29 to 0.69 for large galaxies with different redshifts. The values we find (-1.12 in FB10 and -1.11 in FB20) fit within this range.

As the clumps orbit the galactic center, they shed stars. These stars disperse throughout the galaxy and form the chemical thick disk. The clumps not only enrich the surrounding gas with $\alpha$ elements but scatter stars and convert in-plane motion to vertical motion \citep[][]{Bournaud2009}, similar to the scattering process for giant molecular clouds proposed by \citet{Spitzer1953}. \citet{BeraldoeSilva2020} studied the geometric properties of simulated galaxies with and without clumpy star formation, and found that the one without clumpy star formation did not form a geometric thick disk while the one with clumpy star formation did.

In the FB10 simulation, 29.7\% of stars have high $\alpha$ abundances, while in the FB20 simulation, only 23.1\% of the stars have high $\alpha$ abundances. \citet{Cheng2012} found that the fraction of stars with $\alpha$ abundance greater than 0.2 for $R<10$ kpc varies from 12\% to 31\% for different values of $|Z|$.

\section{Conclusion}
\label{sec:conclusion}
In this work, we have explored two Milky Way simulations employing low-feedback from supernova, namely FB10 and FB20 (for $10\%$ and $20\%$ of the $10^{51}$ erg ejected from each supernova). Our main conclusions are:

\begin{itemize}
\item Both the FB10 and FB20 simulations formed clumps, but the lower feedback FB10 simulation forms clumps for longer.
\item The clumps in the FB10 simulation on average have longer lifetimes and reach higher masses than clumps in the FB20 simulation.
\item Clumps in both simulations that reached high-$\alpha$ on average have a higher mean SFR, a longer lifetime, a greater mass, and form closer to the galactic center than clumps that did not.
\item The radial gradients in mean stellar age of the clumps (-1.12 in FB10 and -1.11 in FB20) are consistent with observed gradients.
\item Most high-$\alpha$ clumps (62\% in FB10 and 54\% in FB20) eventually fall into the galactic center, while most of the low-$\alpha$ clumps dissolve after a few Myr.
\end{itemize}

\acknowledgments
{BRG acknowledges support for this research from the National Science Foundation (AST-1908331).
VPD and LBS are supported by STFC Consolidated grant ST/R000786/1. The simulations in this paper were run at the DiRAC Shared Memory Processing system at the University of Cambridge, operated by the COSMOS Project at the Department of Applied Mathematics and Theoretical Physics on behalf of the STFC DiRAC HPC Facility (www.dirac.ac.uk). This equipment was funded by BIS National E-infrastructure capital grant ST/J005673/1, STFC capital grant ST/H008586/1, and STFC DiRAC Operations grant ST/K00333X/1. DiRAC is part of the National E-Infrastructure.
We would like to thank the anonymous referee for useful suggestions that improved the clarity of the manuscript. We thank Tom Quinn for his help with {\sc gasoline}.
}

\bibliographystyle{aasjournal}
\bibliography{ref_og}

\end{document}